\pdfoutput=1

\documentclass[a4paper,11pt]{article}

\usepackage{cite}
\usepackage[scaled]{berasans}
\usepackage[T1]{fontenc}
\usepackage{amsmath}
\usepackage{amsfonts}
\usepackage{amssymb}
\usepackage{graphicx}
\usepackage{feynmp}
\usepackage{ifpdf}
\usepackage{color}
\usepackage[normalem]{ulem}
\usepackage{simplewick}
\usepackage{hyperref}
\newcommand{\be}{\begin{equation}}
\newcommand{\ee}{\end{equation}}
\newcommand{\ba}{\begin{eqnarray}}
\newcommand{\ea}{\end{eqnarray}}
\newcommand{\bef}{\begin{figure}}
\newcommand{\eef}{\end{figure}}

\newcommand{\al}{\alpha}
 
\newcommand{\e}{\epsilon}
\newcommand{\si}{\sigma}
\newcommand{\ze}{\zeta}

\newcommand{\s}{\sigma}

\newcommand{\nab}{\nabla}

\newcommand{\Famn}{F^a_{\mu \nu}}

\newcommand{\tFmn}{\tilde{F}_{\mu \nu}}

\newcommand{\tFaMN}{\tilde{F}^{\mu \nu \,a}}

\newcommand{\cO}{{\cal O}}

\newcommand{\cP}{{\cal P}}
\newcommand{\cL}{{\cal L}}

\newcommand{\Ra}{\Rightarrow}

\ifpdf
\fi

\def\q{{\bf q}}

\def\k{{\bf k}}

\newcommand{\nn}{\nonumber}

\let\origaligned\aligned
\let\endorigaligned\endaligned
\renewenvironment{aligned}{\wlog{runaligned}\!\origaligned}{\endorigaligned}

\let\origgathered\gathered
\let\endoriggathered\endgathered
\renewenvironment{gathered}{\wlog{rungathered}\!\origgathered}{\endoriggathered}

\newcommand{\acr}{\nonumber\\}
\newcommand{\abs}[1]{\left\lvert #1 \right\rvert}
\newcommand{\bx}[0]{{\mathbf x}}
\newcommand{\bk}[0]{{\mathbf k}}
\newcommand{\bq}[0]{{\mathbf q}}
\newcommand{\mcl}[1]{\mathcal{#1}}

\newcommand{\sO}[0]{\mathcal{O}}
\newcommand{\mbf}[1]{\mathbf{#1}}
\newcommand{\fracs}[2]{{\textstyle\frac{#1}{#2}}}

\newcommand{\til}[0]{\tilde}
\newcommand{\ess}[0]{\hspace{0.07em}}
\newcommand{\DN}[0]{} 

\newcommand{\sci}[1]{\times10^{#1}}
\newcommand{\beps}[0]{{\boldsymbol{\epsilon}}}

\DeclareMathOperator{\re}{Re}
\DeclareMathOperator{\im}{Im}

\makeatletter
\newlength{\@lengthoversetctot}
\newlength{\@lengthoversetcop}
\newcommand{\oversetc}[2]{\settowidth{\@lengthoversetctot}{$\displaystyle{}\overset{#1}{#2}{}$}\settowidth{\@lengthoversetcop}{$\displaystyle {}\overset{}{#2}{}$}\addtolength{\@lengthoversetctot}{-\@lengthoversetcop}{}\overset{\mathclap{#1}}{#2}\hspace*{0.5\@lengthoversetctot}}
\makeatother
\newcommand{\settab}[1]{\hspace*{#1}&\hspace*{-#1}}
\newcommand{\vcenterbox}[1]{\raisebox{-0.5\height}{#1}}

\newcommand{\repart}[1]{\re\left[#1\right]}
\newcommand{\impart}[1]{\im\left[#1\right]}
\newcommand{\sigFFdual}[0]{$\sigma F\til F$ }

\begin{document}

\begin{center}
{\fontsize{22}{24} \fontfamily{\sfdefault}\selectfont On the validity of the perturbative description of axions during inflation} \\[0.5cm]

\large{Ricardo Z. Ferreira$^{\rm a}$, Jonathan Ganc$^{\rm a}$, Jorge Nore\~na$^{\rm b}$, and Martin S. Sloth$^{\rm a}$ }
\\[0.5cm]

\small{
\textit{$^{\rm a}$ {\it  CP$^3$-Origins, Center for Cosmology and Particle Physics Phenomenology}\\
{\it  University of Southern Denmark, Campusvej 55, 5230 Odense M, Denmark}}}

\vspace{.2cm}

\small{
\textit{$^{\rm b}$ University of Geneva, Department of Theoretical Physics and Center for Astroparticle Physics (CAP)\\24 quai E.~Ansermet, 1211 Geneva 4, Switzerland}}

\vspace{.2cm}

\end{center}

\vspace{.8cm}

\hrule \vspace{0.4cm}
\noindent {{\bfseries Abstract} \vspace{0.1cm} \\ 
Axions play a central role in many realizations of large field models of inflation and in recent alternative mechanisms for generating primordial tensor modes in small field models. If these axions couple to gauge fields, the coupling produces a tachyonic instability that leads to an exponential enhancement of the gauge fields, which in turn can decay into observable scalar or tensor curvature perturbations. Thus, a fully self-consistent treatment of axions during inflation is important, and in this work we discuss the perturbative constraints on axions coupled to gauge fields. We show how the recent proposal of generating tensor modes through these alternative mechanisms is in tension with perturbation theory in the in-in formalism. Interestingly, we point out that the constraints are parametrically weaker than one would estimate based on naive power counting of propagators of the gauge field. In the case of non-Abelian gauge fields, we derive new constraints on the size of the gauge coupling, which apply also in certain models of natural large field inflation, such as alignment mechanisms.
}\vspace{0.1cm}
\\ 
\noindent
\hrule

\allowdisplaybreaks

\section{Introduction}
\label{sec:introduction}

It is generally believed that a stage of primordial inflation in the early universe is required in order to resolve the shortcomings of the standard big bang model. The paradigm of primordial inflation has been tested by a number of observations, and so far no inconsistencies with the data have been found, supporting the general principles of the inflationary paradigm \cite{Ade:2015lrj}. On the other hand, the inflationary model space is huge, and we need more experimental input in order to discriminate among the proposed scenarios. One of the most touted prospects is the possibility of primordial tensor modes. Primordial tensor modes are considered to be a smoking gun of inflation, since their amplitude is commonly believed to be directly related to the energy scale of inflation \cite{Starobinsky:1979ty}, and no known alternatives to inflation are expected to lead to an observable level of primordial tensor modes \cite{Gasperini:1992em, Alexander:2000xv, Khoury:2001wf, Creminelli:2010ba}. This has motivated an intensive, ongoing global effort to search for the imprint of primordial tensor modes in the CMB \cite{Ade:2014afa,Ade:2014xna,Baumann:2008aq,Fraisse:2011xz,Benson:2014qhw,Bouchet:2011ck,Matsumura:2013aja}. 

Assuming we detect primordial tensor modes, many results in the literature suggest that axions may have played a central role in their generation. If we believe that the tensors were created by vacuum fluctuations during inflation -- such that their amplitude is related to the energy scale of inflation -- then the Lyth bound tells us that inflation needs to be a large field model, with trans-Planckian field excursions \cite{Lyth:1996im}. However, this makes the inflaton potential sensitive to non-renormalizable Planck-supressed operators unless it is protected by a weakly broken shift-symmetry. The axion, being the pseudo Nambu Goldstone Boson of the broken shift symmetry, is then a natural inflaton in large field models of inflation\footnote{Apart from axions softly broken shift symmetries can also be realized with exponential or power law potentials if the underlying group is non-compact \cite{Burgess:2014tja, Csaki:2014bua}.} \cite{Freese:1990rb, Kim:2004rp,Dimopoulos:2005ac,Silverstein:2008sg,McAllister:2008hb,Kaloper:2008fb, Kaloper:2011jz}.
Alternatively, we can break with the above logic and posit that the tensor modes are not vacuum fluctuations, i.e. that we have some form of \textit{synthetic tensors}. In this case also, the most developed scenarios use axions. Since axions couple to gauge fields with a CP violating axial coupling, a slowly rolling axion during inflation will trigger a resonant production of gauge fields, which can subsequently source tensor modes. It has been speculated that this could lead to an observable tensor-to-scalar ratio, such that the observed tensor modes would not be related to the energy scale of inflation, but instead the microphysics of the resonance\footnote{We review these models in Appendix \ref{app:models}, and the full mode function is given in eq. (\ref{eq:ex-mode_functions}).} \cite{Anber:2006xt, Anber:2009ua, Barnaby:2010vf, Barnaby:2011vw, Sorbo:2011rz, Linde:2012bt, Barnaby:2012xt, Meerburg:2012id, Shiraishi:2013kxa, Cook:2013xea, Mukohyama:2014gba, Bartolo:2014hwa, Ferreira:2014zia, Obata:2014loa, Adshead:2015pva, Fujita:2015iga, Peloso:2015dsa, Bartolo:2015dga, Eccles:2015ipa, Cheng:2015oqa, Namba:2015gja}.  

So if we observe tensor modes, no matter how we view it, axions will be at the locus of the theories suggested to produce them. This motivates a careful understanding of axion cosmology in the very early universe.

The rich phenomenology of the axions is triggered by the presence of the \textit{axial coupling} with gauge fields of the form
\begin{equation}
{\cal L} \supset-  \frac{\al \si}{4 f} \Famn \tFaMN, \nn
\ee
where $f$ is the axion's decay constant, $\Famn$ is the field strength tensor, $\tFmn= 1/2 \, \e^{\mu \nu \al \beta} F_{\al \beta}$ its dual tensor and $\al$ is a dimensionless parameter associated with the axial charge of the axion. 
The presence of the axial coupling triggers an instability in one polarization of the gauge field ($A_+$) propagator such that when a given mode exits the horizon it gets exponentially enhanced as
\begin{equation}\label{eq:intro-enhancement}
  \re[A_+] \propto e^{\pi \xi} \qquad \text{and} \qquad  \im[A_+] \propto e^{- \pi \xi},
\end{equation}
where $\xi= \alpha \dot{\s} /(2 f H)$ is the strength of the exponential enhancement, $H$ is the Hubble constant during inflation, and $\alpha$ is some coefficient related to the axial charge of the axion; note that the field is classicalized at horizon crossing since its real part is exponentially larger than the imaginary part. At the level of perturbations, the axial coupling also induces a new interaction between gauge fields and the scalar curvature perturbation $\zeta$ given by
\begin{align}\label{Lint1}
  {\cal L}_\text{int} = -2 \xi  \, \zeta \, \vec{A} \cdot \left( \nab \times \vec{A} \right) 
\end{align}
which is parametrically stronger than the gravitational interaction and is independent of the role of the axion during inflation \cite{Ferreira:2014zia}. It was shown that if the axion is identified with the inflaton, the combination of gauge field production and large couplings with gravity naturally generates large non-Gaussianities and leads to the constraint $\xi \lesssim 3$ \cite{Barnaby:2010vf}. Subsequently in \cite{Ferreira:2014zia} it was then demonstrated that this constraint also applies if the axion is not identified with the inflaton, unless the axion decays very quickly during inflation, due to the universal nature of the of the coupling in eq (\ref{Lint1}).

Perhaps the most interesting phenomenology triggered by the axial coupling is the possibility of producing tensor modes via the decay of gauge fields \cite{Sorbo:2011rz, Barnaby:2012xt, Mukohyama:2014gba}, i.e. when the diagram in Fig. \ref{fig:enhancement-Abel-1loop} is much larger the tree level result in Fig. \ref{fig:enhancement-tree}. Generally, the coupling between gauge fields and tensor modes is parametrically smaller than the scalar coupling, and so one would expect large tensor modes to be associated with large non-Guassianity \cite{Ferreira:2014zia}. However, if the axion decays during inflation \cite{Mukohyama:2014gba} the non-Gaussianity generated at horizon crossing is erased and the constraint on $\xi$ is relaxed. The efficiency of the mechanism, in generating large tensors, requires some tunning in the decay of the axion \cite{Ferreira:2014zia} but it would represent an alternative for generating primordial tensor modes which would break the connection between the amplitude of the tensor mode power spectrum and the energy scale of inflation.

\begin{figure}
  \centering
  \begin{minipage}[t]{0.45\linewidth}
    \begin{align*}
    \vcenterbox{\includegraphics{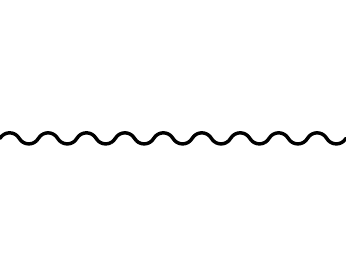}}
  \end{align*}
  \caption{Gauge field 2-point function $\left< AA \right>$}
  \label{fig:bckrx-tree}
  \end{minipage}
  \hspace*{0.01\linewidth}
  \begin{minipage}[t]{0.45\linewidth}
    \begin{align*}
    \vcenterbox{\includegraphics{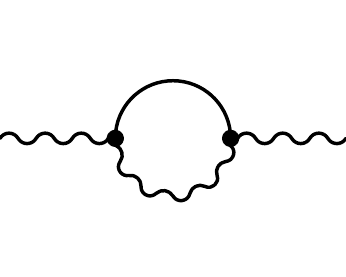}}
  \end{align*}
  \caption{ 1-loop correction to $\left< AA \right>$}
  \label{fig:bckrx-Abel-1loop}
  \end{minipage}
\end{figure}

Since the amplitude of the gauge field is large in this scenario, one has to be concerned about the breakdown of perturbation theory. Indeed, the scenario has some troubling indications. Consider the diagram in Fig. \ref{fig:bckrx-Abel-1loop} with the solid line representing the curvature perturbation $\zeta$. This diagram will appear as a loop in any correlator where a field couples to the gauge field, even if only radiatively, as in the two-point function of $\zeta$. Unless the loop diagram is parametrically suppressed with respect to the tree level Fig. \ref{fig:bckrx-tree}, one must sum an infinite number of diagrams to calculate the correlator, since higher loops of $\zeta$ become increasingly larger. 

\begin{figure}
  \centering
  \begin{minipage}[t]{0.45\linewidth}
    \begin{align*}
    \vcenterbox{\includegraphics{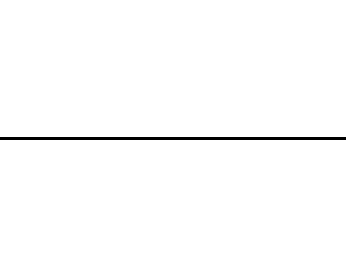}}
  \end{align*}
  \caption{ Scalar (or tensor) 2-point function $\left< \zeta \zeta \right>$ ($\left<h h\right>$)}  \label{fig:enhancement-tree}
  \end{minipage}
  \hspace*{0.01\linewidth}
  \begin{minipage}[t]{0.45\linewidth}
  \begin{align}
    \vcenterbox{\includegraphics{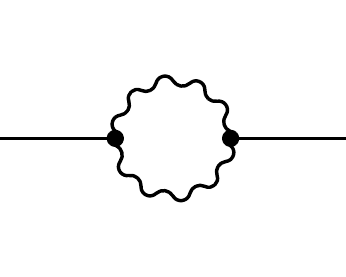}}\nonumber
  \end{align}
  \caption{1-loop correction to $\left< \zeta \zeta \right>$ (or $\left< h h \right>$) } 
  \label{fig:enhancement-Abel-1loop}
  \end{minipage}
\end{figure}

In order to estimate the size of this loop diagram, one may think that it is enough to assign a factor of $\langle A A\rangle \sim e^{2\pi\xi}$ to each propagator of the gauge field. Indeed, the real part of the mode function is exponentially larger than the imaginary part, as apparent in (\ref{eq:intro-enhancement}), and one thus expects it to dominate in any diagram. This would lead us to conclude that the size of the 1-loop diagram relative to the tree level diagram is $e^{4\pi\xi}\xi^2 \cP_\zeta/\ell$, where $\ell\sim 100$ is a loop factor, $\xi^2$ comes from the vertex (\ref{eq:zetAA_itcn}) and $\cP_\zeta$ comes from the scalar propagator. A similar argument tells us that the tensor-mode enhancement, i.e. Fig. \ref{fig:enhancement-Abel-1loop}, is given by a factor $e^{4\pi\xi}\xi^2 \cP_h/\ell$. Thus, it would appear that any enhanced tensor spectrum would immediately imply a complete perturbative breakdown, an immediate and hard no-go for axial coupling synthetic tensors. In fact, demanding perturbativity would seem to yield a very strong constraint on the size of $\xi$, limiting $\xi\lesssim 1.9$.

In this paper, we consider these perturbative questions more carefully. 
As we will see in \ref{sec:counting}, the diagram in Fig \ref{fig:enhancement-Abel-1loop} is ``special'' and, in fact, the naive counting mentioned above does not work for corrections to the $\left<A A\right>$ diagram or for most diagrams. We show how higher loops require an interference between gauge fields at different vertices so that, taking only real parts of the gauge field mode function gives exactly zero. 

In section \ref{sec:perturbative}, we explore the implications of this counting scheme. Parametrically we find the following constraints on $\xi$
\be
 10^{-2} \xi^2 e^{2 \pi \xi} \cP_s \lesssim 1 \qquad \Ra \qquad \xi \lesssim 3.5
\ee
although the constraint from an explicit 1-loop calculation is slightly weaker ($\xi \lesssim 4.3$) due to smaller loop factors.
Though not as strong as those suggested by the naive counting arguments, these constraints leave little space for generating synthetic tensors via axions. Later in the section, we explore constraints on more general models involving axions that couple to gauge fields during inflation, and we look at the scenario where the axion couples to a non-Abelian gauge field, where we find constraints on the gauge coupling strength $g$. We also find that our calculations imply a lower bound for the axion decay constant in the KNP mechanism for natural inflation \cite{Kim:2004rp}. Note that more stringent bounds on $f$ were derived in \cite{Linde:2012bt} for small scales. We compare our parametric estimates with an explicit calculation of the loop diagram, the details of which are presented in Appendix \ref{sec:details-1-loop}. We also review the axial coupling model in Appendix \ref{app:models}.

While we work primarily with the axial coupling model,  we expect this parametric counting to be applicable to other situations where the real part of the mode function is enhanced.

\section{Proper counting in the in-in formalism}
\label{sec:counting}
 
As we have highlighted in the example of the axial coupling discussed above, a naive counting of the parametric enhancement of a diagram in the in-in formalism gives too large of an estimate when considering resonantly produced, classicalized fields.
In this section, we explain how to achieve the correct counting, using rules that account for the cancellations that invalidates the naive estimates. This allows us to properly estimate the perturbative regime.

In the {\it in-in formalism} the expectation value of a given equal-time operator is given by\footnote{The corresponding diagrammatic Feynman style rules for {\it in-in} calculations and non equilibrium problems are given in \cite{Giddings:2010ui}.}  \cite{Calzetta:1986ey}
\be
\left< \Omega \left| \cO_\q(t) \cdots \cO_\k(t) \right| \Omega \right> = \left< 0 \left| \tilde{T} e^{i \int_{-\infty}^t dt' H_\text{int} (t') } \cO_\q(t) \cdots \cO_\k(t) T e^{-i \int_{-\infty}^t dt' H_I (t') } \right| 0\right>, 
\ee
where $\left|\Omega \right>$ represents the vacuum of the interacting theory and $\left|0\right>$ of the free theory, $H_\text{int}(t)$ is the interacting Hamiltonian and $\cO$ is some quantum operator. In the standard perturbative approach, we expand the exponentials in a power series, which can in turn be rearranged into a convenient form\cite{Weinberg:2005vy} \small
\begin{equation}\label{eq:in-in-cmmtr}
  \left< \Omega \left| \cO_\q(t) \cdots \cO_\k(t) \right| \Omega \right> =  \sum_{n=0}^\infty i^n \int_{-\infty}^t dt_n \cdots \int_{-\infty}^{t_2} dt_1 \left< 0 \left| \left[ H_\text{int} (t_1),\cdots, \left[ H_\text{int}(t_n), \cO_\q(t) \cdots \cO_\k(t)  \right] \right] \right| 0\right>.
\end{equation}
\normalsize
This is an exact expression, given as a sum over diagrams with $n \in \{0,1,\ldots\}$ vertices, which then must be integrated over time. For the perturbative expansion to make sense the sum must converge so that diagrams with a larger number of vertices must be suppressed.

A given diagram corresponds to a particular $n$ in \eqref{eq:in-in-cmmtr} along with a particular pattern of contractions among the operators; for example, the 1-loop diagram in Fig. \ref{fig:bckrx-Abel-1loop} corresponds to $n=2$ along with certain contractions, some of which are shown in \eqref{eq:bckrx-Abel-contractions}. In principle, any such set of contractions can be expanded as a sum of terms, each of which contains either the real or the imaginary part of the mode function of each operator. The ``naive'' way of counting the enhancement is to assume there is a term containing a real part from each resonantly-enhanced operator. The problem with this idea is that, because of the commutator nature of the in-in formalism, such a term is often identically zero. Thus, the key goal of this section is to identify the largest term, i.e. to \textbf{identify the non-zero term with the largest number of real parts of resonantly-enhanced operators.}

To identify what real/imaginary combinations are allowed, we must examine the expectation value on the RHS of (\ref{eq:in-in-cmmtr}) with the commutators. As usual, we use Wick's theorem to find these expectation values, where a contraction of an operator $\phi(\bk,\eta)$ is given by\footnote{To simplify the notation we do not differentiate between operators on the left side, and the mode functions after Wick contractions on the right side, but it should be clear from the context which one is which.}
\begin{align}
  \big<\cdots\contraction{}{\phi}{{}_1\cdots}{\phi}\phi_1\cdots\phi_2\cdots\big>
    = (2\pi)^3 \phi_1\phi^*_2\, \delta(\bk_1+\bk_2) \, \big<\,\cdots\,\big>\,,
\end{align}
where the subscript indicates the momentum and time dependence $\phi_1 \equiv \phi (\k_1,t_1)$.
Let us extend this to the case of commutators; in particular, let us suppose that these two operators now lie on different sides of a commutator and that the other mode functions $\psi_1, \psi_2$ which can be any operator, include $\phi$ or a product of $\phi$ or other operators either contract with an operator outside the commutator or are strictly real
\begin{align}
  \big<\cdots\big[\contraction{}{\phi}{{}_1\psi_1,}{\phi}
       \phi_1&\psi_1,\phi_2\psi_2\big]\cdots\big>
    {}= {}(2\pi)^3 \, \delta(\bk_1+\bk_2) \, 
     2i \left(-\repart{\phi_1}\impart{\phi_2} + \impart{\phi_1}\repart{\phi_2}\right)
         \big<\,\cdots\psi_1\psi_2\cdots\,\big>\,.
\end{align}
Based on this, let us make the following simple but important observation:
\begin{align}\label{eq:observation}
  \parbox{\dimexpr\textwidth - 6em\relax}{\textit{To be non-zero, every in-in commutator must have a contraction between two operators, one on each side of the commutator, for which at least one of the mode functions must be imaginary.}}
\end{align}
This can be seen as a requirement of quantum interference between the two (independent) phases of the given mode functions, which cannot happen in the case of classicalized fields where the mode functions are essentially real.

In the axial coupling scenario, we see in \eqref{eq:Aplus_mode_fcn} that the mode function has the resonant behavior\footnote{The time integrals are dominated by the time at which the resonance/classicalization happens, which in the \sigFFdual scenario corresponds to the time of horizon crossing.}
\begin{align}\label{eq:A_resonant}
  \re[A] \propto e^{z( \xi)} \qquad \text{and} \qquad  \im[A] \propto e^{-z( \xi)}\,,
\end{align}
so that each imaginary part cancels the enhancement from one real part. As an simple example, consider
\begin{align}
  \big<\big[ A_1 A_2, A_3 A_4\big] \big>;
\end{align}
the naive parametric enhancement would be to assume we had four real parts, i.e. an enhancement of $e^{4z(\xi)}$, but since we must have one imaginary part, the actual enhancement is $e^{2z(\xi)}$. We now see why naively assigning the size of the power spectrum to each propagator is not accurate. 

We point out that it is not a coincidence that in axially coupled models, the imaginary part of the mode function is the inverse of the real part. It will generally be true because the Wronskian condition $A A'^* - A^* A' = 2i (\im[A] \re[A'] - \re[A] \im[A'])= i$ must hold at all times for a canonically normalized scalar field, so that $\im[A] \re[A]$ will parametrically be of order unity. 

Now let us explore the implications of this observation more generally for the calculation of loops in the $\sigma F\til F$ model involving $\zeta$ and $A$. We write the interaction Hamiltonian schematically as
\begin{equation} \label{eq:intH}
  H_\text{int} \sim \int d^3 x\, \lambda \, \zeta A^2\,,
\end{equation}
so that the correlator in \eqref{eq:in-in-cmmtr}, neglecting Fourier integrals as well as the distinction between $A$ and $A'$, will generally be of the form
\begin{align}
\label{eq:commutator-for-specific-vertex}
  \big< 0 \big| \overbrace{\big[ \zeta A^2 ,\cdots, \big[ \zeta A^2, \zeta^\alpha A^\beta \big] \big]}
      ^{n\text{ commutators}} \big| 0\big>.
\end{align}
Note, first of all, that $\alpha+n$, $\beta$ must be even because the operators must contract in pairs. By our observation \eqref{eq:observation}, we must have $n$ pairs of operators with one imaginary part. However, we can select at most $(\alpha+n)/2$ of these from among the pairs of $\zeta$ so that, if $n>(\alpha+n)/2$ we must have $n-(\alpha+n)/2=(n-\alpha)/2$ pairs of $A$, and thus
\begin{align*}
  \textit{the commutator \eqref{eq:commutator-for-specific-vertex} will have at least }
  \!\max\left\{0,\frac{n-\alpha}{2}\right\}\! \textit{ imaginary mode functions of }A \,.
\end{align*}
We now see why the naive counting works for Fig. \ref{fig:enhancement-Abel-1loop}: the diagram has $\alpha=n=2$ and thus does not need any imaginary parts for $A$. On the other hand, we see why the naive counting fails for Fig. \ref{fig:bckrx-Abel-1loop}: the diagram has $\alpha=0$ but $n=2$ and thus must have at least one imaginary mode function for $A$. Note that even for the case of $\left<\zeta \zeta \right>$ in Fig. \ref{fig:enhancement-Abel-1loop}, \textit{all} higher order loops will follow the more-suppressed counting.

We can now state the size of the parametric enhancements of higher loop diagrams involving this type of vertex. Since additional vertices must come in pairs, consider adding a pair of vertices to a diagram: $\big[ \zeta A^2 ,\big[ \zeta A^2, \cdots \big] \big]$. By the same counting argument as before, at least one of the gauge fields must be imaginary. We have already expressed in \eqref{eq:A_resonant} the resonant behavior of $A$. The curvature perturbation is not resonant so around horizon crossing
\begin{equation}
  \re[\ze] \simeq \im[\ze] \propto \cP^{1/2}\,,
\end{equation}
where\footnote{Note that the same would remain true if $\zeta$ was instead isocurvature or tensor perturbation, in latter case we would have instead $\re[h] \simeq \im[h] \propto H/M_p$ at horizon crossing.} $\cP= 2.2 \times 10^{-9}$. Therefore, the proper counting in the in-in formalism would be 
\begin{align}
\label{sec:parameric-est}
  \left.\frac{\left< A_q A_k \right>_\text{(n+1)-loop}}
      {\left< A_q A_k \right>_\text{n-loop}}\right\rvert_{\zeta A^2}
   = \left.\frac{\left< \zeta_q \zeta_k \right>_\text{(n+1)-loop}}
        {\left<  \zeta_q \zeta_k \right>_\text{n-loop (n>1)}}\right\rvert_{\zeta A^2}
   \sim \frac{\lambda^2}{\ell} \re[A]^3 \im[A] \im[\zeta] \re[\zeta] \sim   \frac{\lambda^2}{\ell} \frac{\Lambda^2}{M^2_p} e^{2 z(\xi)}\,.
\end{align}
Note that we have also included a loop suppression factor of $1/\ell\approx 10^{-2}$. 

It is interesting to note in these setups that, when considering correlators not of the classicalized (i.e. resonant) field, there is a regime where the calculation remains under perturbative control even when the 1-loop correction is significantly larger than the tree level. However, these are special diagrams and, for example, perturbativity still breaks down when the 1-loop diagram of the classicalized field becomes larger than the tree level.

\section{Perturbative Constraints}
\label{sec:perturbative}

We now apply the parametric counting derived in the previous section to the case of an axial coupling between an axion-like particle $\si$ and a $U(1)$ or $SU(N)$ gauge field, as reviewed in Appendix \ref{app:models}.  To begin with, let us consider the case of a $U(1)$ gauge field and compare the two diagrams in Figs. \ref{fig:bckrx-tree} and \ref{fig:bckrx-Abel-1loop}. From the arguments of section \ref{sec:counting}, we can use \eqref{sec:parameric-est} to find the parameteric size of the relative amplitude of the diagrams
\be\label{eq:loop_suppres} 
\frac{\left\langle A_k A_q \right\rangle_\text{1-loop}}{\left< A_k A_q \right>_\text{tree}}  \approx \frac{\xi^2}{l} e^{2 \pi \xi} \cP_s, 
\ee
where we use the strength $\xi$ of the coupling from the vertex (\ref{eq:zetAA_itcn}). To stay in the perturbative regime, this quantity must be less than unity. Thus, we obtain the parametric constraint
\be
\mathrm{\it Perturbativity:} \qquad \xi \lesssim 3.5.
 \label{eq:perturb_axion}
\ee

So far, we have used fairly general arguments for the size of additional loops but have not calculated any diagrams. Thus, it behooves us to find a scenario where we can compare the size of a diagram with that of a similar one but with an additional loop of the type shown in Fig. \ref{fig:bckrx-Abel-1loop}. This will give us a sanity check on our results. Specifically, we will calculate the 1-loop correction to $\left<A A\right>$ in the case of an Abelian gauge field. 

The dominant tree level calculation follows immediately from canonical quantization
\begin{align}
  &\left< A_l(\bk,\eta) A_l(\bk',\eta)\right>_\text{tree} 
     = (2\pi)^3 \abs{A_+(k,\eta)}^2 \delta(\bk+\bk')\,, \acr
  &\left< A_l'(\bk,\eta) A_l'(\bk',\eta)\right>_\text{tree} 
      = (2\pi)^3 \abs{A_+'(k,\eta)}^2 \delta(\bk+\bk')\,,
\end{align}
where the mode functions $A_+(k,\eta)$ are given by \eqref{eq:ex-mode_functions}. The first correction is, in fact, the diagram in Fig. \ref{fig:bckrx-Abel-1loop}. Since this is not a ``special diagram'', as in the case in Fig. \ref{fig:enhancement-Abel-1loop}, we expect its amplitude to be given by \eqref{eq:loop_suppres}.

We explicitly performed the calculation of the 1-loop correction to $\left< AA \right>$ (details are presented in Appendix \ref{sec:details-1-loop}). In Fig. \ref{fig:numer_prtbtvty}, we present the ratio of the tree-level and 1-loop terms for both $\left< A^2\right>$ and $\left< A'^2\right>$. 
We see that the 1-loop result becomes as large as the tree-level result at $\xi \geq 4.4$ for both $\left< A^2\right>$ and $\left< A'^2\right>$. This is slightly weaker than our parametric estimate \eqref{eq:loop_suppres} due to smaller loop factors coming from the contractions of polarization vectors.

\begin{figure}[ht]
  \centering
  \includegraphics[width=0.5\textwidth]{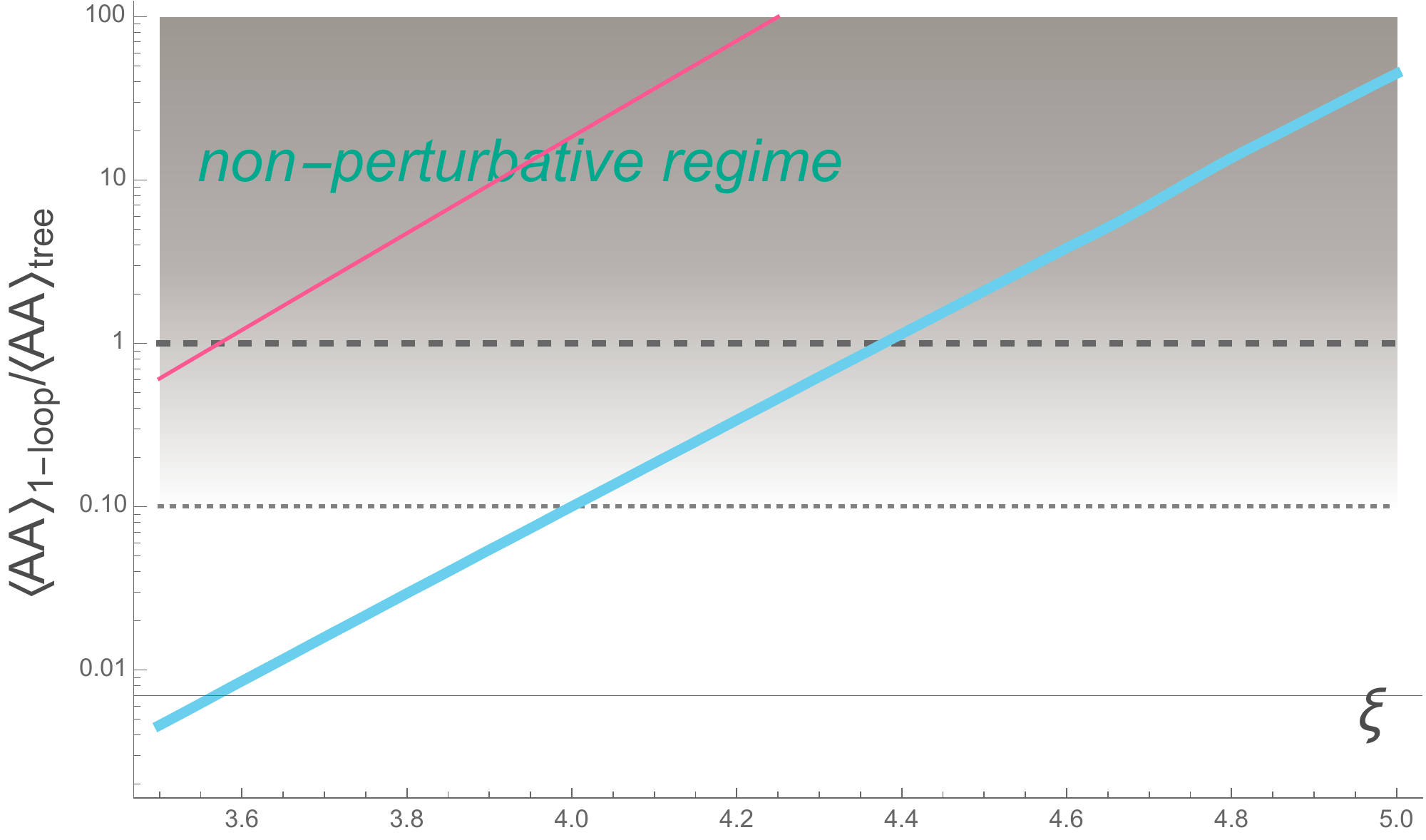}
  \caption{The thick, blue line shows the ratio of the numerical 1-loop calculation of $\left\langle AA \right\rangle$ divided by the tree level calculation (the result for $\left\langle A'A' \right\rangle$ is very slightly larger). The thin, red line shows the parametric estimate using \ref{sec:parameric-est}, which is considerably more restrictive than the result from the numerical 1-loop calculation. \ref{sec:details-1-loop}.}
\label{fig:numer_prtbtvty}
\end{figure}

We can also consider another bound that is similar in spirit, namely from demanding the validity of the effective description used to treat the model. If the resonance is too strong, the fluctuations of the axion $\delta\sigma$ are enhanced through loop corrections as in Fig. \ref{fig:enhancement-Abel-1loop} and can possibly grow larger than the period of the potential \hbox{$\delta\sigma \gtrsim f$}. If this happens, higher-order interactions between the axion and the gauge fields become equally important as the ones that are usually studied, e.g. in the Lagrangian \eqref{eq:sigFFdual-Lgrn}. In that regime, non-linear effects become important, such as the damping of the fluctuations themselves due to the periodicity of the potential. Using the calculation of the loop correction to the axion perturbations presented in \cite{Barnaby:2010vf}, we can estimate the condition this imposes on the parameters of the model
\begin{equation}
	\label{eq:small_sigma}
	\mathrm{ \it Small\ axion\ fluctuations:}\quad3\times 10^{-4} \cP_s^2 \frac{e^{4\pi\xi}}{\alpha^2\xi^4} \lesssim 1.
\end{equation}
For $\alpha=\mcl{O}(1)$, this implies $\xi<4.3$. Note that this result is sensitive to $\alpha$, the axial charge of the axion; in the standard model, for example, $\alpha=1/137$ and the bound is $\xi>3.4$, which would be in tension with the generation of synthetic tensor modes.
In Fig. \ref{fig:constraints} we plot the constraints on synthetic tensors. On the left-hand panel, we plot in green the region where the production of tensor modes is incompatible with current bounds from the BICEP2/Keck array \cite{Array:2015xqh}. In blue, we plot the region where the tensor modes sourced by the vacuum account for more than a fifth of the total tensor mode production. We show the results for a wide range of values of the slow-roll parameter $\epsilon$. We see that there is still a region of parameter space which is not excluded by these requirements. This region requires large values of $\xi$ which would be excluded by current bounds on non-Gaussianity \cite{Ferreira:2014zia}, though it is possible to tune the decay of the axion in such a way that this non-Gaussianity disappears \cite{Mukohyama:2014gba,Ferreira:2014zia}. In the middle panel we overlay in dark (light) orange the constraint from demanding the ratio of the 1-loop to tree-level diagram to be 1 (0.1). The window for the mechanism to work bcomes significantly reduced. Finally, in the right-hand panel, we plot in dark (light) yellow the region where the oscillations of the axion are larger than the period of the potential, where \eqref{eq:small_sigma} does not hold, for $\alpha=1$ ($\alpha=1/137$). The constraint gets stronger for weaker coupling. In particular, for electromagnetic-like couplings there would be no parameter space to generate consistently gravitational waves larger than the vacuum. We have been conservative and considered the bound on $\xi$ from our explicit calculation of the one-loop diagram. We see that, even though these bounds depend on the details of the calculations and are sensitive to coefficients of order one, there is little space for this mechanism to work.

Let us mention briefly that this constraint also puts bounds on models which require large values of $\xi$, such as models that dissipate the inflaton's kinetic energy through the production of vectors as in \cite{Anber:2009ua}. These models require a strong coupling between vectors and the inflaton in order for the energy dissipation to be efficient. In that reference the authors use $\xi \approx 20$. In light of what we discussed above, this model is expected to produce large non-Gaussianities and tensor modes, while also being in the non-perturbative regime.

\begin{figure}
	\centering
	\includegraphics[width=5cm]{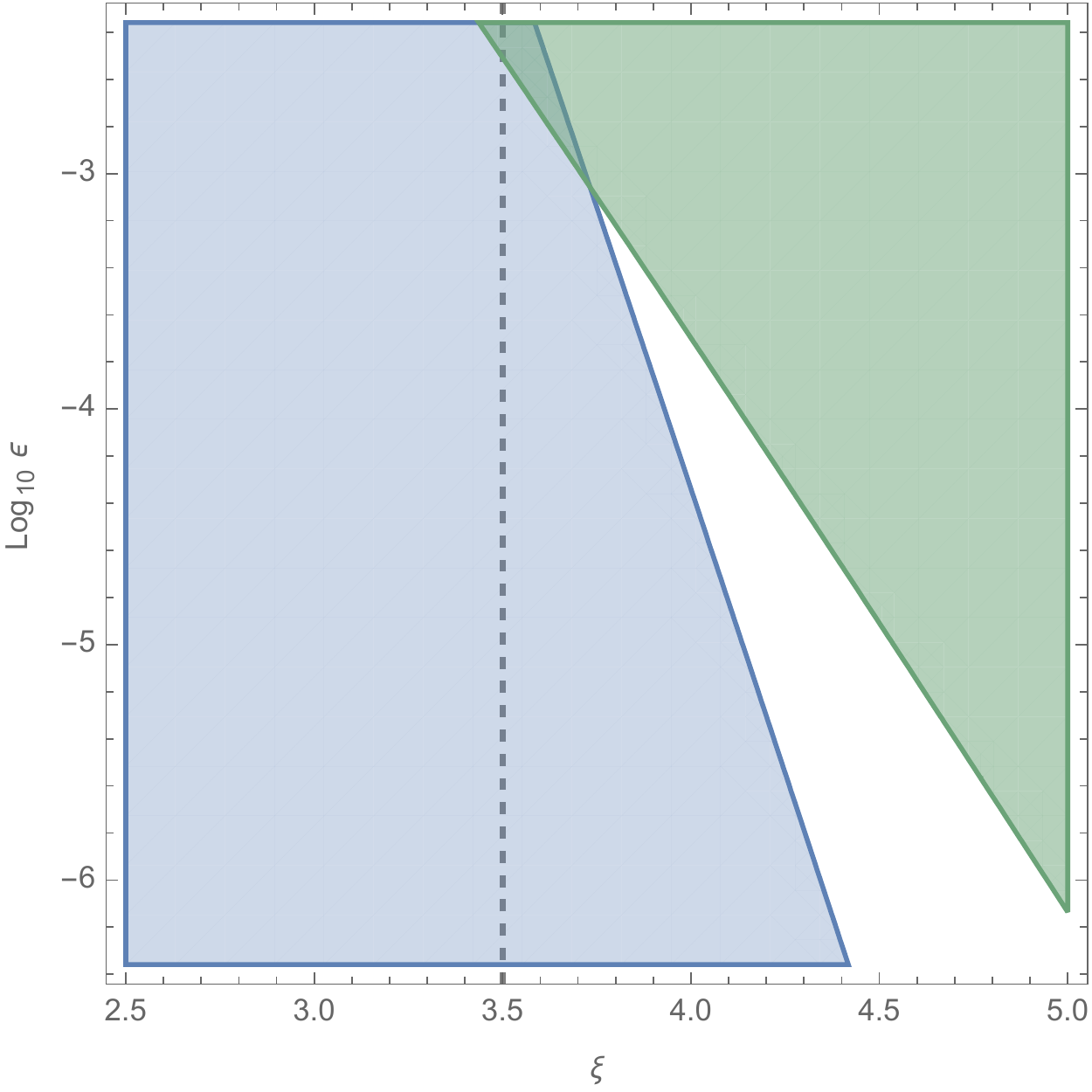}
	\includegraphics[width=5cm]{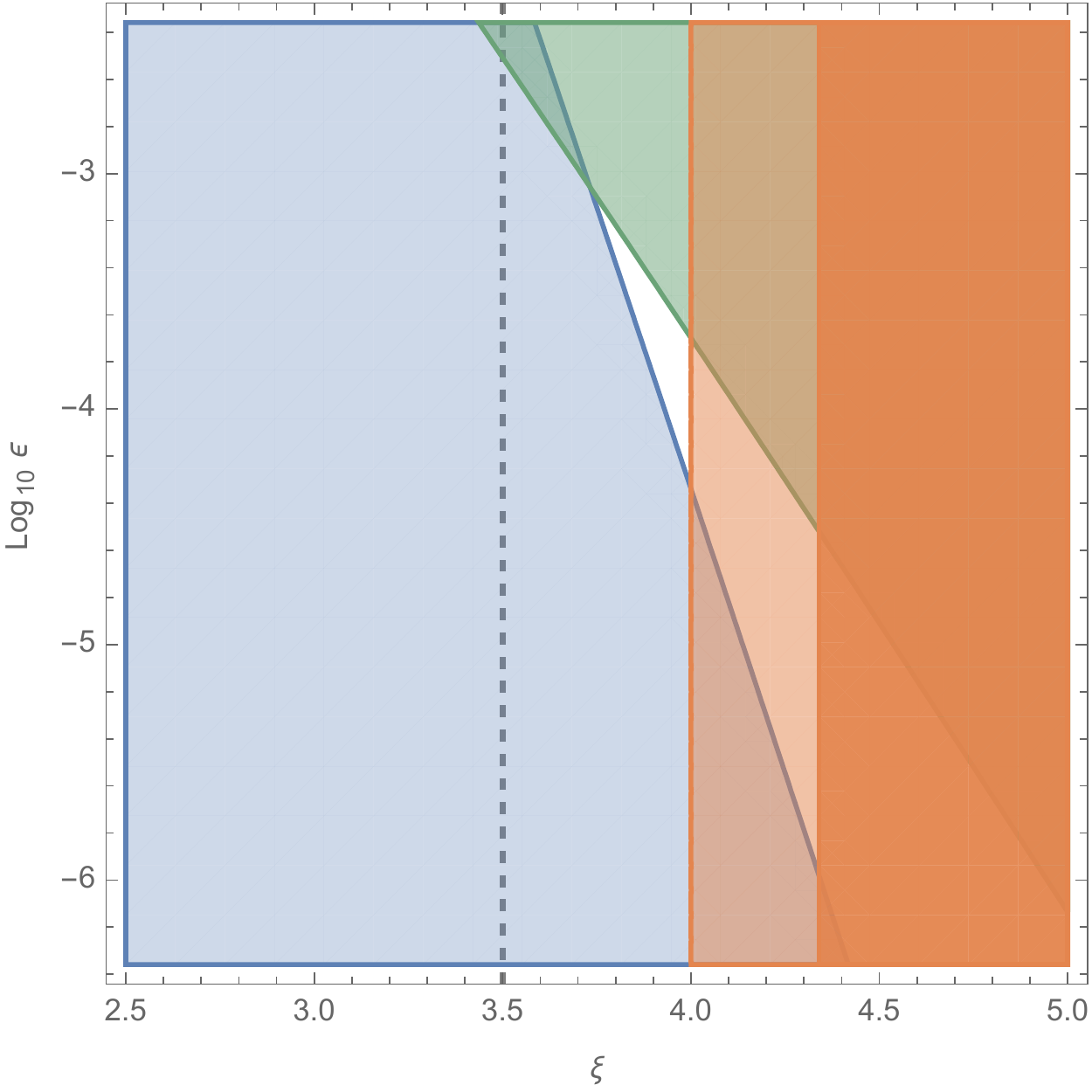}
	\includegraphics[width=5cm]{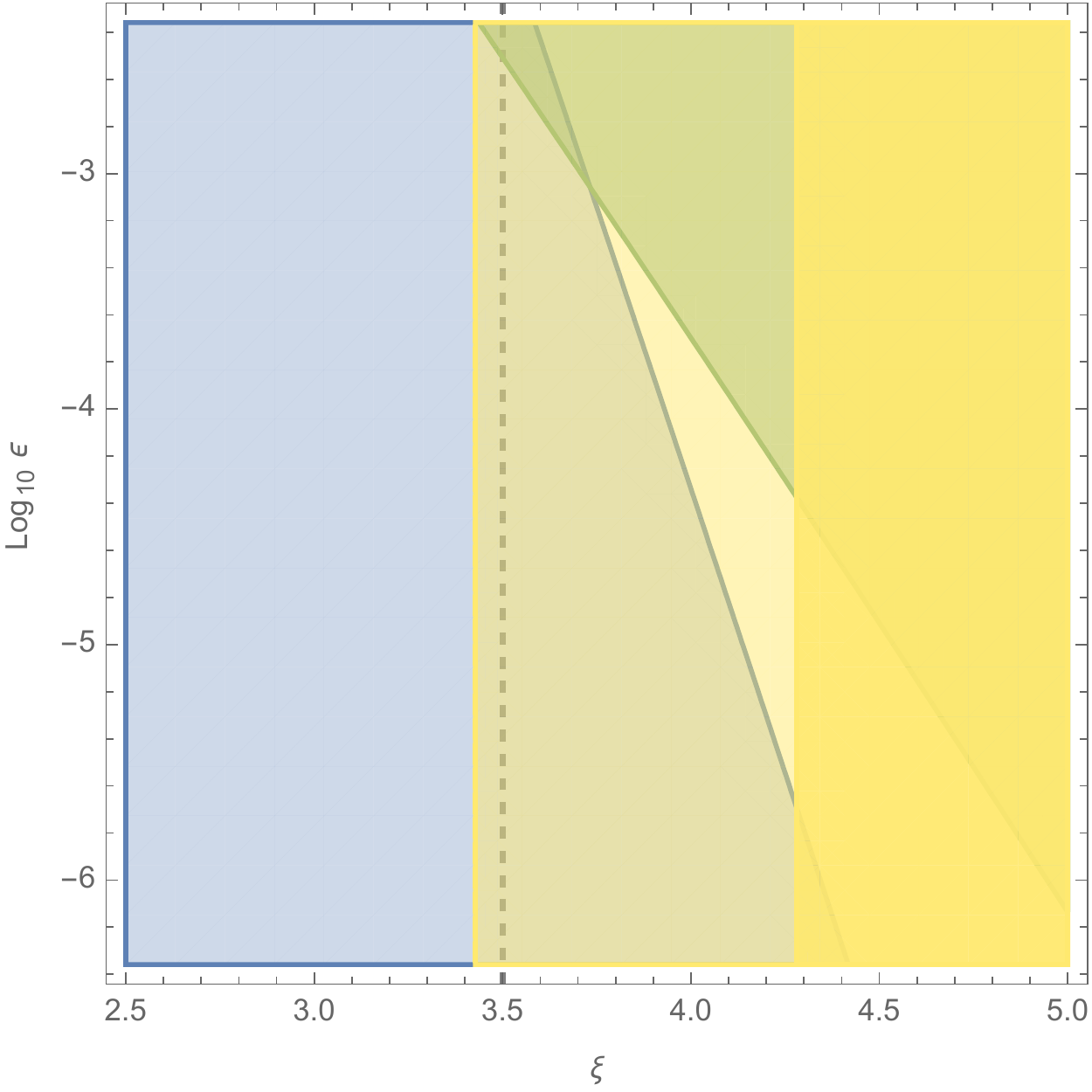}
	\caption{Shown in blue is the region where tensor modes sourced by the vacuum are more than a fifth of the total amplitude. Shown in green is the region excluded by current constraints on the tensor-to-scalar ratio \cite{Array:2015xqh}, both for a wide range of values of $\epsilon$. The dashed gray line shows the perturbativity constraint from assuming $\ell=\mcl O(10^2)$. Middle Panel: We overlay constraints from demanding perturbativity (i.e. a convergent loop expansion); the dark (light) orange region is excluded by requiring that the calculated ratio of the 1-loop to tree level diagram be 1 (0.1). Right panel: We overlay the EFT constraint \eqref{eq:small_sigma}; the dark (light) yellow region is excluded by the requirement of small fluctuations of the axion, setting $\alpha = 1 \; (1/137)$.}  \label{fig:constraints}
\end{figure}

We can also consider the parametric enhancements of loops from the non-Abelian vertices \eqref{eq:AAA_itcn}-\eqref{eq:zetAAA_itcn}, shown in Fig. \ref{fig:non-Abelian-loops}. Starting with the leftmost diagram, which is cubic in the gauge fields, we proceed as before, noting that we have six gauge fields but that two must be imaginary, giving us
\begin{align} 
   \left.\frac{\left< A_q A_k \right>_\text{(n+1)-loop}}
      {\left< A_q A_k \right>_\text{n-loop}}\right\rvert_{A^3}
     \sim \frac{g^2 \left(f^{abc}\right)^2}{\ell} \re[A]^4 \im[A]^2  
     \sim \frac{g^2N^3}{\ell} e^{2\pi\xi} \,, \label{NAconstrA3}
\end{align}
where we have used the fact that $\left(f^{abc}\right)^2\propto N^3$ (e.g. \cite{pesk}, where $N$ is from the group $SU(N)$). The middle diagram, quartic in $A$, gives
\begin{align} 
   \left.\frac{\left< A_q A_k \right>_\text{(n+1)-loop}}
      {\left< A_q A_k \right>_\text{n-loop}}\right\rvert_{A^4} 
    \sim \frac{g^4 \left(f^{abc}\right)^4}{\ell^2} \re[A]^6 \im[A]^2  
    \sim \frac{g^4N^6}{\ell^2} e^{4\pi\xi} \,, \label{NAconstrA4}
 \end{align}
and the right diagram, using $\zeta A^3$ vertices, yields
\begin{align} 
   \left.\frac{\left< A_q A_k \right>_\text{(n+1)-loop}}
      {\left< A_q A_k \right>_\text{n-loop}}\right\rvert_{\zeta A^3} 
    \sim \frac{\xi^2 g^2 \left(f^{abc}\right)^2}{\ell^2} \re[A]^5 \im[A] \re[\zeta] \im[\zeta] 
     \sim \frac{\xi^2 g^2N^3}{\ell^2} e^{4\pi\xi} \mcl P_\zeta \,. \label{NAconstrZetaA3}
\end{align}
In the non-Abelian case, the constraint from the $\zeta A^2$ vertex also gets a factor equal to the number of generators in the gauge group, so the constraint becomes $\xi^2N^2\mcl P_\zeta e^{2\pi\xi}/\ell$. In itself, this has little impact on predictions about non-Gaussianity or synthetic tensor production from this vertex because these effects will also be enhanced by $N^2$. Note also that the enhancements in (\ref{NAconstrA4}) and (\ref{NAconstrZetaA3}) are just a combination of the Abelian (\ref{eq:loop_suppres}) and cubic self-interaction (\ref{NAconstrA3}) enhancements and therefore, ultimately, those are the ones which should be smaller than one. Therefore, for the non-Abelian case the constraints are:
\ba  \label{gConst}
\text{\it Perturbativity (non-Abelian case):} \qquad  &\xi^2N^2\mcl P_\zeta e^{2\pi\xi}/\ell <1 & \nn \\
&g^2 e^{4 \pi \xi} N^3/\ell< 1.& 
\ea

\newcommand{\nonabelianloopwidth}[0]{100}
\newcommand{\nonabelianloopheight}[0]{80}
\newlength{\nonabelianminipagewidth}

\begin{figure}
  \setlength{\nonabelianminipagewidth}{0.3\linewidth}
  \centering
  \begin{minipage}[t]{\nonabelianminipagewidth}
    \begin{align*}
    \vcenterbox{\includegraphics{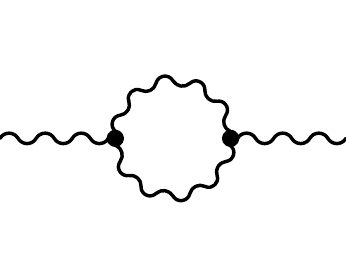}}
  \end{align*}
  \end{minipage}
  \hspace*{0.01\linewidth}
  \begin{minipage}[t]{\nonabelianminipagewidth}
  \begin{align*}
    \vcenterbox{\includegraphics{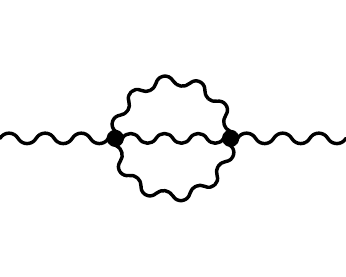}}
  \end{align*}
  \end{minipage}
  \begin{minipage}[t]{\nonabelianminipagewidth}
  \begin{align*}
    \vcenterbox{\includegraphics{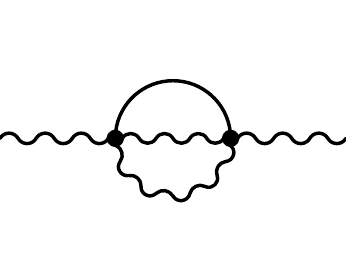}}
  \end{align*}
  \end{minipage}
  \caption{The non-Abelian loops, from the non-Abelian vertices \eqref{eq:AAA_itcn}, \eqref{eq:AAAA_itcn}, \eqref{eq:zetAAA_itcn}.}
  \label{fig:non-Abelian-loops}
\end{figure}

\subsection{Consequences for natural inflation}

In the context of UV completions of natural inflation \cite{Dimopoulos:2005ac,Silverstein:2008sg,McAllister:2008hb,Kaloper:2008fb, Kaloper:2011jz} consistency arguments have been put forward which translate into an upper bound on the axion decay constants \cite{Banks:2003sx, Baumann:2006cd, Conlon:2011qp, Conlon:2012tz, Kaloper:2015jcz, Palti:2015xra, ArkaniHamed:2006dz, Brown:2015iha}. But, as already noted by several authors \cite{Barnaby:2010vf, Barnaby:2011vw, Barnaby:2011qe}, the phenomenology of the axial coupling and the consequent non-Gaussian constraints on the parameter $\xi$ also provides interesting lower bounds. In \cite{Ferreira:2014zia} the constraints were generalized to all axions during inflation by identifying the universal coupling to gravity generated by the axial coupling. Interestingly, the results derived here show that the phenomenologically allowed regions previously associated with large observable non-Gaussianity is not under perturbative control. 

More specifically, in scenarios such as the KNP mechanism, where several axions couple to the same gauge groups, the mass basis (which contains the light/inflaton direction) is typically not the same as the coupling basis to gauge fields. Therefore, axions with small decay constants become potentially dangerous. For example, consider the original setup,  where two axions $\phi_1$, $\phi_2$ are coupled to two different gauge groups $F, G$ as \cite{Kim:2004rp} 
\ba
\cL \supset &&-\Lambda_1 \left[1+ \cos \left( \frac{\phi_1}{f} + \frac{\phi_2}{g_1} \right) \right] - \Lambda_2 \left[1+ \cos \left( \frac{\phi_1}{f} + \frac{\phi_2}{g_2} \right) \right] - \nn \\ && - \frac{1}{4} \left( \frac{\alpha_1 \phi_1}{f}  + \frac{\alpha_2 \phi_2}{g_1} \right) F_{\mu \nu}^a \til{F}^{a\, \mu \nu}  - \frac{1}{4} \left( \frac{\beta_1 \phi_1}{f}  + \frac{\beta_2 \phi_2}{g_2} \right) G_{\mu \nu}^a \til{G}^{a\, \mu \nu}\,.
\ea
By rewriting the Lagrangian in the mass basis, the light direction $\si = f g_1/\sqrt{f^2 +g_1^2} (\phi_2/f-\phi_1/g_1)$ (which would appear if $g_1\simeq g_2$) would then couple to the gauge fields as 
\be
- \frac{\si}{4 \sqrt{f^2 +g_1^2}} \left[ (\alpha_1- \alpha_2) F_{\mu \nu}^a \til{F}^{a\, \mu \nu}  +(\beta_1- \beta_2) G_{\mu \nu}^a \til{G}^{a\, \mu \nu} \right].
\ee
Therefore, the requirement $\xi \lesssim 3.5$ would imply
\be
\sqrt{f^2 +g_1^2} \gtrsim 4 \times 10^{-2} (\alpha_1-\alpha_2) M_p,  
\ee
where we have fixed $\epsilon= \dot \si^2/(2 M_p^2 H^2)=0.04$. As we can see from the previous expression, decay constants smaller than the GUT scale may lead to the perturbative breakdown of inflation.

Similar reasoning can be done for axion monodromy. In the frame where the gauge fields are canonically normalized, the time dependence of the coupling to the Chern-Simons term should satisfy the above constraints on $\xi$ in order for perturbativity to be guaranteed in the computation of correlators involving $\zeta$, $h_i$ or $A_\mu$.

We can also consider the implications on the gauge group coupling constant from eq. (\ref{gConst}). For example, for decay constants of the order of the GUT scale such that $\xi \simeq \cO (1)$, the constraint would imply $g \lesssim 10^{-2}$.

\section{Conclusions}

Axions have a rich phenomenology in early universe cosmology. Among other things they are candidates for large field inflation models and may also provide a mechanism for boosting the production of tensor modes through their axial coupling with gauge fields. This coupling term is a total derivative unless the axion is rolling. Even if the axion is rendered subdominant in the energy budget of the universe, its rolling triggers an instability in the gauge fields which then, in turn, act as a source of metric perturbations. As a consequence, tensor perturbations may be produced with an amplitude larger than the ones due to vacuum production, which generically requires $\xi \gtrsim 3.5$, where $\xi$ is the the strength of the instability \cite{Mukohyama:2014gba,Ferreira:2014zia}. However, they will also generate scalar perturbations of the metric \cite{Ferreira:2014zia} with
 large non-Gaussianity; imposing that this must be compatible with current bounds requires $\xi \lesssim 3$. This non-Gaussianity can, however, be erased if the axion decays during inflation. Though this requires some tuning, it would seem to allow the generation of synthetic tensor modes. This possibility would break the standard connection between the energy scale of inflation and the tensor-to-scalar ratio.

In this paper, we have discussed that there may be another set of constraints that rule out or, at least, severely limit these scenarios: the resonant production of gauge fields when $\xi > 1$ brings with it a risk that perturbation theory breaks down. In this work we investigated how to determine the perturbative region in inflation in the presence of fields which are resonantly enhanced and so, quickly classicalized. We provided a way to parametrically estimate the relative size higher loop diagrams in the presence of such resonances. Interestingly, we found that in the in-in formalism, as particularly apparent when it is written in the commutator form \eqref{eq:in-in-cmmtr}, the terms that would show the most resonant behavior typically cancel out, so that the enhancement of higher order loops is not pronounced as one would naively expect from a simple counting of propagators. 

Based on this counting we derived perturbative constraints on axions coupled axially to gauge fields. Parametrically we found that in order to ensure perturbativity of the loop expansion, the strength of the interaction should satisfy $\xi \lesssim 3.5$, although an explicit 1-loop calculation led to a weaker constraint ($\xi \lesssim 4.4$). This constraint, jointly with the EFT constraint from requiring $\delta \phi<f$ which becomes the strongest one for weaker coupling, leaves little space for the mechanism to generate tensor modes larger than the vacuum (in a controllable regime). We note that, though this bound is weaker than that implied from constraints on non-Gaussianity \cite{Barnaby:2010vf} and black hole formation \cite{Linde:2012bt}, it must hold also at the time production so that it is independent of the evolution of the universe after production, making it more robust. We also note that in the case of a coupling with a non-Abelian gauge field, the additional vertices imply an additional constraint; in particular, one must demand that $ g^2 e^{4 \pi \xi} N^3/\ell< 1$, where $g$ is the gauge coupling and $\ell\sim10^2$ is a loop factor. 

One should however keep in mind that even in the parameter regions where perturbativity breaks down, we cannot rule out synthetic tensor modes, but rather can only state that the perturbative calculations on which the predictions are based are inapplicable.

Finally, we stress that for models of inflation where axions play an important role (e.g. UV completed models of natural inflation) these constraints might be relevant and require, in some cases, either small axial charges or $f>10^{16}$ GeV  scale, where $f$ is the axion decay constant, and typically implies $g \lesssim 10^{-2}$ for the gauge coupling.

There are several avenues for future work. Since this work challenges the mechanism producing tensor modes through axial couplings, it would be interesting to do a detailed study of constraints on other possibilities for generating synthetic tensors, such as those mentioned in \cite{Senatore:2011sp, Carney:2012pk, Mirbabayi:2014jqa, Biagetti:2013kwa}. If model builders wanted more precise constraints on $\xi$, one could examine the vertices and in-in calculation more closely and derive more aggressive bounds on the size of the loop factors, as well as to bound the factors (such as $\xi$'s) that come from the time integrations. 

\section*{Acknowledgments}

We would like to thank Cliff Burgess, Daniel Baumann, Guido D'Amico and Nemanja Kaloper for enlightening discussions. We also thank Marco Peloso, Lorenzo Sorbo and Caner Unal for pointing out a numerical inaccuracy. RZF, JG and MSS would like to thank the Lundbeck foundation for financial support. CP3-Origins is partially funded by the Danish National Research Foundation, grant number DNRF90. J.N. is supported by the Swiss National Science Foundation (SNSF), project ``The non-Gaussian Universe'' (project number: 200021140236).
\appendix

\section{An overview of the axially coupled model}\label{app:models}

Consider a model with a axion-like particle $\sigma$ and $SU(N)$ gauge field $A_\mu^a$, with the action
\begin{align}\label{eq:sigFFdual-Lgrn}
  \mcl{L} &= -\frac{1}{2} (\partial_\mu \sigma)^2
     -\frac{1}{4} F^a_{\mu\nu}  F^{a\ess \mu\nu} - \frac{\alpha\sigma}{4f} F_{\mu\nu}^a \til F^{a\ess\mu\nu}\,,
\end{align}
where $f$ is the axion decay constant of $\sigma$; we use the Hodge dual $\tilde F_{\mu\nu} \equiv \frac{1}{2} \sqrt{-\|g_{\mu\nu}\|} \epsilon_{\mu\nu\alpha\beta} F^{\alpha\beta}$, where the Levi-Civita tensor $\epsilon_{\mu\nu\alpha\beta}$ is totally antisymmetric in its indices and $\epsilon_{0123}=1$; and $\alpha$ is the coupling constant of the parity violating coupling. 

The field strength tensor of the non-Abelian $A$ is defined as usual:
\begin{align}
  F^a_{\mu\nu} = A^a_{\nu,\mu} - A^a_{\mu,\nu} + g f^{abc} A^b_\mu A^c_\nu\,.
\end{align}
where $f^{abc}$ is the structure constant associated with the $SU(N)$ generators $\{T^a\}$, i.e. $[T^a, T^b] = i f^{abc} T^c$; we can choose $f^{abc}$ to be completely antisymmetric.  We will typically work in Coulomb gauge, where $A_0=A_{i,i}=0$. (Note that we sum over repeated color indices and repeated vector indices, regardless of upper vs. lower placement.)

For constant $\sigma$, the third term in the Lagrangian is a total derivative. Thus, the  effective strength of the interaction is sensitive only to the time dependence of the background part of $\sigma(\bx,t)=\sigma(t) + \delta\sigma(\bx,t)$, i.e. to $\dot\sigma_0$, and it is convenient to define
\begin{align}
  \xi\equiv \alpha\dot\sigma/2fH\,.
\end{align}

\subsection{Mode functions for $A$}

The quadratic, Coulomb-gauge action for $A^a_{\mu}$, as given by \eqref{eq:sigFFdual-Lgrn}, is
\begin{align}
  S_2 &= \int d^3x d\eta \left[ -\frac{a^4}{2} (\partial_\mu \delta\sigma)^2
     + \frac{1}{2} \left(A'_i A'_i - A_{i,j} A_{i,j}\right)
        - a \xi H \epsilon^{ijk} A^a_i A^a_{k,j}\right],
\end{align}
which gives the equations of motion $\mbf A'' + k^2 \mbf A + 2 i \xi a H \, \bk \times\mbf A = 0$.

To decouple the spatial degrees of freedom of the gauge field in the equation of motion, it is useful to decompose the gauge fields in terms of (complex) circular polarization vectors $\epsilon_{\pm\ess i}(\bk)$, so that (assuming Coulomb gauge)
\begin{align}\label{eq:decompose_A_clscl}
  A_i(\bk,\eta) = \sum_\lambda A_\lambda(\bk,\eta) \epsilon_{\lambda\ess i}(\bk)\,,
\end{align}
where we choose $\epsilon_{\pm\ess i}(\bk)$ to satisfy the following conditions
\begin{align}
  &\epsilon_{\lambda\ess i}^*(\bk) \epsilon_{\lambda'\ess i}(\bk) = \delta_{\lambda\lambda'} \,, && k_i \epsilon_{\pm\ess i}(\bk) = 0\,,
   && \bk\times\beps_{\pm\ess}(\bk) = \mp ik\beps_\pm(\bk)\,,
   && \beps_\pm(-\bk) = \beps_\pm^*(\bk)\,.
\end{align}
We can choose
\begin{align}
  \beps_\pm(\hat k) &= 
   \frac{1}{\sqrt2} \left(\cos\theta\cos\phi \mp i\sin\phi, \cos\theta\sin\phi \pm i\cos\theta,-\sin\theta\right).
\end{align}

Then, the equations of motion for $A_\pm$ become
\begin{align}\label{eq:eqn_for_A}
  \left(\frac{\partial^2}{\partial \eta^2} + k^2 \pm \frac{2k\xi}{\eta}\right)A_\pm(k,\eta) = 0\,.
\end{align}
We see that the $A_+$ mode has a tachyonic instability when $-k\eta<2\xi$, leading to resonant production of gauge field particles, and therefore is our primary object of interest.

The Bunch-Davies mode function for $A_+$ is the solution to \eqref{eq:eqn_for_A} which approaches $e^{-ik\eta}$ for modes deep within the horizon, i.e. as $-k\eta\to \infty$. If we also demand that the quantized field $A$ obey the standard quantization condition $[A_i(\bx),\Pi^A_j(\mbf y)]=i \int \frac{d^3k}{(2\pi)^3} e^{i \bk\cdot(\bx - \mbf y)} (\delta_{ij} - \delta_{jm} k_i k_m / k^2)$, we get the further condition on the Wronskian $W_\lambda(k,\eta) \equiv A_\lambda(k,\eta) A_\lambda^{\prime *}(k,\eta) - A_\lambda^*(k,\eta) A_\lambda'(k,\eta) = i$. This condition, which must always hold, is important to keep in mind because it highlights the fact that the real part of a mode function times its conjugate is generally of order unity.

We ultimately find that the properly normalized mode functions for $A_\pm$ are \cite{Anber:2009ua}:
\begin{align}\label{eq:ex-mode_functions}
  A_\pm = \frac{1}{\sqrt{2k}} H_0^+(\pm\xi,-k\eta),
\end{align}
where $H_0^+$ is an irregular Coulomb function (see \cite{NIST:DLMF} for properties of $H_0^+$).

We are interested in the exponentially enhanced cases, where $\xi$ is large\footnote{More accurately, when $2\xi/(\!-\!k\eta)\gg1$.} and where we can approximate
\begin{align}\label{eq:APl_xi_lg}
  A_+(\eta,k) &{}\approx \sqrt{\frac{-\eta}{2}}
      \left[2 e^{\pi\xi} \pi^{-1/2} \operatorname{K}_1\left(\sqrt{-8k\xi\eta}\right)
        + i e^{-\pi\xi} \pi^{1/2} \operatorname{I}_1(\sqrt{-8k\xi\eta})\right]\,,
\end{align}
using the modified Bessel functions I$_\ell$, K$_\ell$. The effects are dominated by the horizon crossing regime, where $-k\eta\approx1$ so $\sqrt{-8k\eta}\gg1$. In this regime, we can approximate (\cite{Anber:2009ua})
\begin{align}\label{eq:Aplus_mode_fcn}
  &A_+(\eta,k) \approx \frac{1}{\sqrt{2k}} \left(\frac{-k\eta}{2\xi}\right)^{1/4}
    e^{\pi\xi - 2\sqrt{-2\xi k \eta}}
    + \frac{i}{\sqrt{2k}} \left(\frac{-k\eta}{2^5 \xi}\right)^{1/4}
    e^{-\pi\xi + 2\sqrt{-2\xi k \eta}}\,,\acr
  &A_+'(\eta,k) \approx \sqrt{\frac{k}{2}} \left(\frac{2\xi}{-k\eta}\right)^{1/4}
    e^{\pi\xi - 2\sqrt{-2\xi k \eta}}
    - i \sqrt{\frac{k}{2}} \left(\frac{\xi}{- 2^3k \eta}\right)^{1/4}
    e^{-\pi\xi + 2\sqrt{-2\xi k \eta}}\,.
\end{align}
This approximation is valid in the regime $(8\xi)^{-1} \lesssim -k\eta\lesssim 2\xi$. One can easily see that the real parts of the mode functions are exponentially and parametrically enhanced while the imaginary parts are likewise suppressed.

\subsection{Scalar and vector interactions}

Let us ignore tensor couplings. As pointed out in \cite{Ferreira:2014zia}, we can most easily find the Lagrangian in terms of the
gauge-invariant curvature perturbation $\zeta$ by using spatially flat gauge, where the spatial metric (including tensor perturbations $\gamma$) is $g_{ij} = a^2 \delta_{ij}$. In this gauge, we can write the perturbation $\delta\sigma$ in the axion field as
\begin{align}
  \delta\sigma = \frac{\dot\sigma}{H} \left(\zeta + \mcl S_{\sigma\phi} \right),
\end{align}
where $\mcl S_{\sigma\phi}$ is the gauge-invariant isocurvature between the inflaton $\phi$ and axion $\sigma$ (which can be taken to be zero if $\sigma$ is the inflaton). Then we can effectively read off the relevant interactions between scalar perturbations and gauge fields:; we will also include the non-Abelian gauge term self-couplings from the kinetic term\footnote{Note that there is no $\zeta A^4$ term because there is no $A^4$ term in $F^a_{\mu\nu} \til F^{a\ess\mu\nu}$. To see this, let us write down the would-be $A^4$ term
\begin{align}\label{eq:expl-FFdual}
  F^a_{\mu\nu} \til F^{a\ess\mu\nu} \supset
    \fracs{1}{2} g^2 (-\|g_{\mu\nu}\|)^{-1/2}\epsilon^{\mu\nu\alpha\beta} f^{abc} f^{ade} A^b_\mu A^c_\nu A^d_\alpha A^e_\beta\,.
\end{align}
However (see e.g. \cite{Macfarlane:1968ab}),
\begin{align}
  f^{abc} f^{ade} 
    = \frac{2}{N} \left(\delta^{bd} \delta^{ce} - \delta^{be} \delta^{cd}\right) 
      + \left(d^{abd} d^{ace} - d^{acd} d^{abe}\right),
\end{align}
where $d^{abc}$ can be taken to be completely symmetric. Substituting this into \eqref{eq:expl-FFdual} and multiplying out the terms, one can verify by swapping indices that each term equals minus itself and thus vanishes.
}
\begin{subequations}
\label{eq:sclr-itcns}
\begin{align}
  \label{eq:zetAA_itcn}
 & \cL_{\zeta A^2} = -2 \xi\DN \epsilon^{ijk} \zeta A_{i}^{a\ess\prime} A^a_{j,k} ,&\\
  \label{eq:AAA_itcn}
 & \cL_{A^3} =  -g f^{abc} \left(A^a_{j,i} A^b_i A^c_j
     +  \xi H \epsilon^{ijk} A^a_i A^b_j A^c_k\right),&\\
  \label{eq:AAAA_itcn}
  &\cL_{A^4} =  - \fracs{1}{4} g^2 f^{abc} f^{ade} A^b_i A^c_j A^d_i A^e_j\,,&
     \\
  \label{eq:zetAAA_itcn}
  &\cL_{\zeta A^3} =  \fracs{1}{3} \xi g f^{abc} \epsilon^{ijk} \zeta' A^a_i A^b_j A^c_k \,;&
\end{align}
\end{subequations}
the first vertex is for the Abelian case and the other three arise in the non-Abelian case. As emphasized in \cite{Ferreira:2014zia}, this \sout{these} interaction are present even if $\sigma$ is not the inflaton.

\section{Details of the 1-loop correction to $\left<AA\right>$ calculation}
\label{sec:details-1-loop}

Let us calculate the 1-loop correction to $\left<AA\right>$, corresponding to Fig. \ref{fig:bckrx-Abel-1loop}, which has two $\zeta A^2$ vertices given by \eqref{eq:sclr-itcns} The corresponding in-in expression is
\begin{align}\label{eq:eq:bckrx-Abel-1loop-in-in}
  \settab{3em}\left< A_i(\bk,\eta) A_i(\bk',\eta) \right>\acr
    &= i^2 \int_{-\infty}^\eta d\eta_B \int_{-\infty}^{\eta_B} d\eta_A
     \left<\left[H(\eta_A), \left[H(\eta_B), 
        A_i(\bk,\eta) A_i(\bk',\eta) \right] \right]\right>\acr
    &\begin{aligned}= 4\xi^2 \int &\prod_{q=1}^6 d^3q_i \times (2\pi)^{-12} \delta(1+2+3) \delta(4+5+6)
      \int_{-\infty}^\eta d\eta_B \int_{-\infty}^{\eta_B} d\eta_A
      \epsilon^{lmn} \epsilon^{pqr} \\
       &{}\times q_{3\ess n} q_{6\ess r}
        \left<\big[\zeta(1) A'_l(2) A_m(3), \big[\zeta(4) A'_p(5) A_q(6),
          A_i(k) A_i(k') \big] \big]\right>.
      \end{aligned}
\end{align}
where $A_i(a)\equiv A_i(\bq_a,\eta_a)$ and where $\eta_a$ is either $\eta_A$ or $\eta_B$, as appropriate for $\bq_a$. Considering that mode functions of $A$ and $\zeta$ have real and imaginary parts, the full result of the above expression has many terms. But for large $\xi$, the dominant terms will be those with the largest number of real parts of mode functions of $A$. 

Using the framework developed in Sec. (\ref{sec:counting}), we know that we must have at least two imaginary mode functions since we have two commutators. Only one of these can be chosen to be from the curvature perturbations $\zeta$ since these contract with each other; thus, we must have at least one imaginary mode function from a gauge-field $A$. Furthermore, this imaginary $A$ must be from one of the $4$ $A$'s in the inner commutator, since each commutator must have a contraction inside it with an imaginary part; we can take the remaining $A$-contractions to be real. We remember that, if the operators $\sO_i^x$ are real or contract with operators outside of the commutator, then
\begin{align}
  \big[\sO_1^L&\contraction{}{\phi}{{}_1\sO_1^R, \sO_2^L}{\phi_2}
     \phi_1\sO_1^R, \sO_2^L\phi_2\sO_2^R\big]
    {}= \big[\contraction{}{\phi}{{}_1, }{\phi} \phi_1, \phi_2\big] 
       \sO_1^L\sO_1^R \sO_2^L\sO_2^R\acr
    &{}= 2 i (2\pi)^3 (-\repart{\phi_1} \impart{\phi_2} + \repart{\phi_2} \impart{\phi_1})\, \delta(1+2) \;
     \sO_2^L\sO_2^R \sO_1^L\sO_1^R\,.
\end{align}

Thus, we can calculate the part of this correlation that is most enhanced by tachyonic instability
\begin{align}\label{eq:bckrx-Abel-contractions}
  \settab{2em}\left<\big[\zeta(1) A'_l(2) A_m(3), \big[\zeta(4) A'_p(5) A_q(6),
     A_i(k) A_i(k') \big] \big]\right>_\text{enhanced}\acr
   =\DN{}&\big[\contraction{}{\zeta}{(1),}{\zeta}\zeta(1),\zeta(4)\big]
      \Big[\begin{aligned}[t]
        \Big(&\big< A'_l(2) A_m(3) \big[\contraction{}{A}{{}'_p(5) A_q(6), }{A}
            A'_p(5) A_q(6), A_i(k) A_i(k') \big] \big>\\
        &{}+ \big< A'_l(2) A_m(3) \big[ A'_p(5)
           \contraction{}{A}{{}_q(6),}{A} A_q(6),
          A_i(k) A_i(k') \big]\big>\Big) + k\leftrightarrow k'\Big]
 \end{aligned}\acr
   =\DN{}&\big[\contraction{}{\zeta}{(1),}{\zeta}\zeta(1),\zeta(4)\big]
      \Big[\begin{aligned}[t]
        \Big(&\big[\contraction{}{A}{{}'_p(5),}{A} A'_p(5) , A_i(k) \big]
          \big< A'_l(2) A_m(3) A_q(6) A_i(k')  \big>\\
        &{}+ \big[\contraction{}{A}{{}_q(6),}{A} A_q(6) , A_i(k) \big]
            \big< A'_l(2) A_m(3) A'_p(5) A_i(k') \big>\Big) + k\leftrightarrow k'\Big]\,.
        \end{aligned}
\end{align}

Plugging into \eqref{eq:eq:bckrx-Abel-1loop-in-in} and simplifying -- using also the identity for the polarization vectors\\ $\abs{\left[\beps(\bk)\times\beps(\bq_3)\right]\cdot\bq_3}^2 = \fracs{1}{4}q_3^2 (1-\hat\bk\cdot\hat\bq_3)^2$ --  we arrive at the expression for 1-loop correction to $\left<AA\right>$
\begin{align}\label{eq:backr_Abel-full}
  \settab{0.5em}\left< A_i(\bk,\eta) A_i(\bk',\eta) \right>_\text{1-loop, enhanced}\acr
    &\begin{aligned}{}= &2^3\mcl \xi^2 \delta(k+k') \repart{A(k,\eta)}
        \int d^3q\, (1-\hat\bk\cdot\hat\bq)^2\, \int_{-\infty}^\eta d\eta_B \int_{-\infty}^{\eta_B} d\eta_A \\
       &{}\times\DN\big(-\repart{\zeta(q',\eta_A)} \impart{\zeta(q',\eta_B)} + \impart{\zeta(q',\eta_A)} \repart{\zeta(q',\eta_B)}\big)_{\bq'=-\bk-\bq}\\
       &{}\times \big(q \repart{A'(k,\eta_A)} \repart{A(q,\eta_A)} + k \repart{A'(q,\eta_A)} \repart{A(k,\eta_A)}\big)\\
       &\begin{aligned}{}\times \Big[&q \,\big(-\repart{A'(k,\eta_B)} \impart{A(k,\eta)} 
               + \repart{A(k,\eta)} \impart{A'(k,\eta_B)} \big) \repart{A(q,\eta_B)}  \\
         &{}+k \,\big(-\repart{A(k,\eta_B)} \impart{A(k,\eta)} 
                  + \repart{A(k,\eta)} \impart{A(k,\eta_B)} \big) \repart{A'(q,\eta_B)}\Big]\,,
       \end{aligned}
     \end{aligned}
\end{align}
where all gauge mode functions $A=A_+$ are from the $+$ polarization. The result for\\ $\langle A_i'(\bk,\eta) A_i'(\bk',\eta) \rangle_\text{1-loop, enhanced}$ is the exactly the same except that $A(k,\eta), A(k',\eta)\to A'(k,\eta), A'(k',\eta)$.

We then numerically integrate the expression \eqref{eq:backr_Abel-full} using the exact mode functions \eqref{eq:ex-mode_functions} for $A$, along with the standard mode function for $\zeta$ (see e.g. \cite{Malda,wein-cosmo}): $\zeta(k,\eta) = \pi\sqrt{2\mcl P_s(k)/k^3} (1+ik \eta) e^{-ik \eta}$, where $\mcl P_s=2\sci{-9}$ is the observed perturbation 2-point function.

The integral \eqref{eq:backr_Abel-full} does not have a IR divergence; in particular, for fixed $k\eta$, there does not need to be a IR cutoff for the $\eta_a$ or $q$ integrals. On the other hand. the mode functions for both $A$ and $\zeta$ are oscillatory inside the horizon (to see this for $A$, simply take \eqref{eq:eqn_for_A} with $-q\eta_a\gg1$), which mimics standard Minkowski-space behavior. In general, we would need to renormalize or regularize away the 1-loop UV divergence. Since the standard, oscillatory behavior begins when $-q\eta_a>2\xi$, we cut off the mode functions $A(-q,\eta_a)$ at this value to approximate an adiabatic subtraction scheme, i.e. we take $A(-q,\eta_a)=0$ when $-q\eta_a>2\xi$.

Using this scheme, we find the results used in Fig. \ref{fig:numer_prtbtvty}.

\bibliographystyle{JHEP}
\bibliography{PerturbativeConstraints}

\end{document}